\documentclass[prl,preprint,superscriptaddress,showpacs,amsmath,amssymb]{revtex4}

\ifx\pdfoutput\undefined
\usepackage[dvips]{graphicx}
\else
\usepackage[pdftex]{graphicx}
\usepackage{epstopdf}
\fi

\setkeys{Gin}{width=.99\columnwidth, height=.75\textheight, keepaspectratio}
\usepackage{dcolumn}
\usepackage{bm}
\usepackage{url}
\usepackage{xspace}

\newenvironment{cfigure}[1][tbp]{\begin{figure}[#1]\centering}{\end{figure}}

\newcommand{\fig}[1]{Fig.~\ref{#1}}

\newcommand{\Fig}[1]{Figure~\ref{#1}}
\newcommand{\Tab}[1]{Table~\ref{#1}}

\newcommand{\ttbar}  {\ensuremath{t\bar{t}}\xspace}

\newcommand{\et}     {\ensuremath{E_{T}}\xspace}

\newcommand{\pt}     {\ensuremath{p_{T}}\xspace}

\newcommand{\chisq}  {\ensuremath{\chi^{2}}\xspace}

\newcommand{\mtop}   {\ensuremath{\mathrm{M}_{\text{top}}}\xspace}
\newcommand{\mreco}  {\ensuremath{\mathrm{m}_{\mathrm{t}}^{\text{reco}}}\xspace}

\newcommand{\cdfii}  {CDF~II\xspace}
\newcommand{\DZero}  {D\O\xspace}

\newcommand{\jes}    {\ensuremath{\mathrm{JES}}\xspace}

\newcommand{\gev}[1] {\ensuremath{#1~\mathrm{GeV}}}

\newcommand{\gevc}[1] {\ensuremath{#1~\mathrm{GeV}/c}}

\newcommand{\gevcc}[1]{\ensuremath{#1~\mathrm{GeV}/c^{2}}}
\newcommand{\gevccnoarg}{\ensuremath{\mathrm{GeV}/c^{2}}}
\newcommand{\invpb}[1]{\ensuremath{#1~\mathrm{pb}^{-1}}}

\newcommand{\sigcunit}[1]{\ensuremath{#1~\sigma_c}\xspace}

\newcommand{\measErr}[2]{\ensuremath{#1 \pm #2}\xspace}
\newcommand{\measAErr}[3]{\ensuremath{#1~^{+#2}_{-#3}}\xspace}

\newcommand{\measAStat}[3]{\ensuremath{#1~^{+#2}_{-#3}~(\text{stat.})}\xspace}

\newcommand{\measAStatSyst}[4]{\ensuremath{#1~^{+#2}_{-#3}~(\text{stat.})\pm #4~(\text{syst.})}\xspace}

\begin{document}

%
%
%

\title{Precision Top Quark Mass Measurement in the 
Lepton + Jets Topology in $p\bar{p}$\ Collisions at 
$\sqrt{s}=1.96$~TeV}
\affiliation{Institute of Physics, Academia Sinica, Taipei, Taiwan 11529, Republic of China} 
\affiliation{Argonne National Laboratory, Argonne, Illinois 60439} 
\affiliation{Institut de Fisica d'Altes Energies, Universitat Autonoma de Barcelona, E-08193, Bellaterra (Barcelona), Spain} 
\affiliation{Baylor University, Waco, Texas  76798} 
\affiliation{Istituto Nazionale di Fisica Nucleare, University of Bologna, I-40127 Bologna, Italy} 
\affiliation{Brandeis University, Waltham, Massachusetts 02254} 
\affiliation{University of California, Davis, Davis, California  95616} 
\affiliation{University of California, Los Angeles, Los Angeles, California  90024} 
\affiliation{University of California, San Diego, La Jolla, California  92093} 
\affiliation{University of California, Santa Barbara, Santa Barbara, California 93106} 
\affiliation{Instituto de Fisica de Cantabria, CSIC-University of Cantabria, 39005 Santander, Spain} 
\affiliation{Carnegie Mellon University, Pittsburgh, PA  15213} 
\affiliation{Enrico Fermi Institute, University of Chicago, Chicago, Illinois 60637} 
\affiliation{Joint Institute for Nuclear Research, RU-141980 Dubna, Russia} 
\affiliation{Duke University, Durham, North Carolina  27708} 
\affiliation{Fermi National Accelerator Laboratory, Batavia, Illinois 60510} 
\affiliation{University of Florida, Gainesville, Florida  32611} 
\affiliation{Laboratori Nazionali di Frascati, Istituto Nazionale di Fisica Nucleare, I-00044 Frascati, Italy} 
\affiliation{University of Geneva, CH-1211 Geneva 4, Switzerland} 
\affiliation{Glasgow University, Glasgow G12 8QQ, United Kingdom} 
\affiliation{Harvard University, Cambridge, Massachusetts 02138} 
\affiliation{Division of High Energy Physics, Department of Physics, University of Helsinki and Helsinki Institute of Physics, FIN-00014, Helsinki, Finland} 
\affiliation{University of Illinois, Urbana, Illinois 61801} 
\affiliation{The Johns Hopkins University, Baltimore, Maryland 21218} 
\affiliation{Institut f\"{u}r Experimentelle Kernphysik, Universit\"{a}t Karlsruhe, 76128 Karlsruhe, Germany} 
\affiliation{High Energy Accelerator Research Organization (KEK), Tsukuba, Ibaraki 305, Japan} 
\affiliation{Center for High Energy Physics: Kyungpook National University, Taegu 702-701; Seoul National University, Seoul 151-742; and SungKyunKwan University, Suwon 440-746; Korea} 
\affiliation{Ernest Orlando Lawrence Berkeley National Laboratory, Berkeley, California 94720} 
\affiliation{University of Liverpool, Liverpool L69 7ZE, United Kingdom} 
\affiliation{University College London, London WC1E 6BT, United Kingdom} 
\affiliation{Massachusetts Institute of Technology, Cambridge, Massachusetts  02139} 
\affiliation{Institute of Particle Physics: McGill University, Montr\'{e}al, Canada H3A~2T8; and University of Toronto, Toronto, Canada M5S~1A7} 
\affiliation{University of Michigan, Ann Arbor, Michigan 48109} 
\affiliation{Michigan State University, East Lansing, Michigan  48824} 
\affiliation{Institution for Theoretical and Experimental Physics, ITEP, Moscow 117259, Russia} 
\affiliation{University of New Mexico, Albuquerque, New Mexico 87131} 
\affiliation{Northwestern University, Evanston, Illinois  60208} 
\affiliation{The Ohio State University, Columbus, Ohio  43210} 
\affiliation{Okayama University, Okayama 700-8530, Japan} 
\affiliation{Osaka City University, Osaka 588, Japan} 
\affiliation{University of Oxford, Oxford OX1 3RH, United Kingdom} 
\affiliation{University of Padova, Istituto Nazionale di Fisica Nucleare, Sezione di Padova-Trento, I-35131 Padova, Italy} 
\affiliation{University of Pennsylvania, Philadelphia, Pennsylvania 19104} 
\affiliation{Istituto Nazionale di Fisica Nucleare Pisa, Universities of Pisa, Siena and Scuola Normale Superiore, I-56127 Pisa, Italy} 
\affiliation{University of Pittsburgh, Pittsburgh, Pennsylvania 15260} 
\affiliation{Purdue University, West Lafayette, Indiana 47907} 
\affiliation{University of Rochester, Rochester, New York 14627} 
\affiliation{The Rockefeller University, New York, New York 10021} 
\affiliation{Istituto Nazionale di Fisica Nucleare, Sezione di Roma 1, University of Rome ``La Sapienza," I-00185 Roma, Italy} 
\affiliation{Rutgers University, Piscataway, New Jersey 08855} 
\affiliation{Texas A\&M University, College Station, Texas 77843} 
\affiliation{Istituto Nazionale di Fisica Nucleare, University of Trieste/\ Udine, Italy} 
\affiliation{University of Tsukuba, Tsukuba, Ibaraki 305, Japan} 
\affiliation{Tufts University, Medford, Massachusetts 02155} 
\affiliation{Waseda University, Tokyo 169, Japan} 
\affiliation{Wayne State University, Detroit, Michigan  48201} 
\affiliation{University of Wisconsin, Madison, Wisconsin 53706} 
\affiliation{Yale University, New Haven, Connecticut 06520} 
\author{A.~Abulencia}
\affiliation{University of Illinois, Urbana, Illinois 61801}
\author{D.~Acosta}
\affiliation{University of Florida, Gainesville, Florida  32611}
\author{J.~Adelman}
\affiliation{Enrico Fermi Institute, University of Chicago, Chicago, Illinois 60637}
\author{T.~Affolder}
\affiliation{University of California, Santa Barbara, Santa Barbara, California 93106}
\author{T.~Akimoto}
\affiliation{University of Tsukuba, Tsukuba, Ibaraki 305, Japan}
\author{M.G.~Albrow}
\affiliation{Fermi National Accelerator Laboratory, Batavia, Illinois 60510}
\author{D.~Ambrose}
\affiliation{Fermi National Accelerator Laboratory, Batavia, Illinois 60510}
\author{S.~Amerio}
\affiliation{University of Padova, Istituto Nazionale di Fisica Nucleare, Sezione di Padova-Trento, I-35131 Padova, Italy}
\author{D.~Amidei}
\affiliation{University of Michigan, Ann Arbor, Michigan 48109}
\author{A.~Anastassov}
\affiliation{Rutgers University, Piscataway, New Jersey 08855}
% ...
\collaboration{CDF Collaboration}
\noaffiliation

\date{\today}

\begin{abstract}
We report two measurements of the top quark mass \mtop\ using the \cdfii 
detector at the Fermilab Tevatron 
in a $\invpb{318}$ data sample observed in the lepton + jets
final state. 
One method uses an event-based likelihood technique
resulting in
$\mtop=\gevcc{\measAStatSyst{173.2}{2.6}{2.4}{3.2}}$\
or $\gevcc{\measAErr{173.2}{4.1}{4.0}}$. 
The second method reconstructs a top quark mass in
each event using the measured invariant mass of the hadronically
decaying $W$ boson to constrain the jet energy scale to obtain
a value for \mtop of
$\gevcc{\measAStatSyst{173.5}{3.7}{3.6}{1.3}}$\ or
$\gevcc{\measAErr{173.5}{3.9}{3.8}}$.  
We take the latter, which is more precise, as our result. 
\end{abstract}


\maketitle

%
%
%

The top quark is the heaviest known elementary particle with a mass
approximately 40 times that of the next-heaviest quark or lepton.
Because of this comparatively large mass, 
top quark studies provide insight into our understanding of mass in
general, and test theories that explain the large range of quark
and lepton masses.
Within the context of the 
Standard Model of particle physics, 
the top quark mass is related to the masses
of the $W$ boson and the Higgs boson, the latter
object being the key to our understanding of the origin of 
mass~\cite{r_electroweakfit}.
Precision measurements of the top quark and $W$ boson masses 
test the consistency of the Standard Model, and in particular the Higgs
mechanism. 
A precision measurement of the top quark mass is therefore 
a main goal of the experiments at the Fermilab Tevatron collider.

In this Letter we present two measurements of the top quark mass in the
lepton + jets decay channel.
We use a sample of \ttbar\ decays corresponding to
\invpb{318} of proton-antiproton collisions at $\sqrt{s}=1.96$\ TeV 
and collected 
using the Collider Detector at Fermilab
(CDF II) between February 2002 and August 2004.
In the lepton + jets channel, \ttbar pair production is followed by
the decay of each top quark to a $W$ boson and a $b$ quark, the
hadronic decay of one $W$\ boson, and the leptonic decay of the other.
This decay channel has the largest branching fraction with good 
signal-to-background, allowing accurate top quark mass measurements.
Events in this channel contain an electron or muon and a neutrino
from the leptonic $W$ boson decay, two quark jets from the hadronic
$W$ boson decay, and two $b$-quark jets.

We select events consistent with this decay topology and analyze
them using two complementary methods.  
The first method uses an
event-by-event likelihood analysis employing the leading order matrix element
for \ttbar production and decay to extract a joint likelihood as a
function of the top quark mass, \mtop.
This technique, known as the ``dynamical likelihood method'' or DLM,
was developed by the CDF collaboration~\cite{Kondo:1988yd} and 
is similar to that used by the \DZero collaboration to make the 
previous most precise measurement of the top quark mass~\cite{Abazov:2004cs}.
The second method, developed by the CDF
collaboration~\cite{r_evidence_prd}, reconstructs a top quark
mass, \mreco, in each event and compares the distribution of \mreco\
with template distributions derived from
model calculations to estimate \mtop.
We have improved this ``template method'' 
by making further use of the fact that the hadronically 
decaying $W$ boson daughters
should form a final state whose invariant mass is consistent with 
the known $W$ boson mass and width.  This allows us to constrain the jet
energy scale, an important uncertainty in the earlier measurements.
These two  methods have a top quark mass
accuracy 30\%\ greater than earlier results, and have 
different statistical and systematic uncertainties.

%
%
The CDF II detector 
\cite{r_cdfDetector}\
is a general-purpose charged and neutral particle
detector located at the
Tevatron collider.
We employ cylindrical coordinates
where $\theta$ and $\phi$ are the polar and azimuthal angles, respectively, 
with respect to the proton
beam, and pseudorapidity is 
$\eta = -\ln\tan(\theta/2)$.  Transverse energy and momentum are 
$E_T= E\sin\theta$ and $p_T=p\sin\theta$, respectively, where $E$ and
$p$ are energy and momentum. 
The detector comprises a solenoidal charged particle spectrometer, 
consisting of 
an eight-layer silicon microstrip detector array and a cylindrical drift
chamber immersed in a 1.4 T magnetic field, a
segmented sampling calorimeter with acceptance
up to pseudorapidity $|\eta|=3.6$,
and a set of charged particle detectors outside the
calorimeter used to identify muon candidates.

Events for these analyses were selected by requiring an
electron or muon candidate with transverse momentum
$\pt>\gevc{20}$ and $|\eta|<1$\ and missing
transverse energy exceeding \gev{20}, corresponding to a
high-energy neutrino candidate.  
The signal-to-background was improved by requiring in each event the
presence of four or more jets with $|\eta|<2.0$. 
To reduce backgrounds further
we required either (a) at least four jets with transverse energy 
$\et>\gev{21}$\ or (b) at least three jets with $\et>\gev{15}$\ and
a fourth jet with $\et>\gev{8}$\ with at least
one jet with $\et>\gev{15}$\ identified as a $b$ quark candidate through 
the presence of 
a displaced vertex within
the jet arising from the decay of the long-lived bottom 
hadron ($b$ tag).
This selection resulted in 165 events that, based on our
background estimates, are primarily \ttbar events.
The methods used to estimate the backgrounds are detailed
in~\cite{r_run2topxsec}.

The DLM analysis uses a 63-event subset of those data defined by
requiring exactly four jets with $\et>\gev{15}$ 
where at least one of the jets has a $b$ tag. 
We have estimated the various sources of background contamination in this 
sample, summarized in \Tab{t:backgroundnumber}, to be
$9.2\pm1.8$\ events.
The template method divides the 165 events into four non-overlapping subsamples with 
different expected \mreco distributions and background levels.
Ordered by decreasing statistical power, the subsamples are
1) events with at least four jets with $\et>\gev{15}$ and one $b$-tagged jet
(``1-tag Tight'' sample with 63 events),
2) events with two or more $b$-tagged jets
(``2-tag'' sample with 25 events),
3) events with a fourth jet with $\gev{8}<\et<\gev{15}$ and one $b$-tagged jet
with $\et>\gev{15}$\ (``1-tag Loose'' sample with 33 events), and
4) events with four jets with $\et>\gev{21}$ and no $b$-tagged jets
(``0-tag'' sample with 44 events).
The estimated background levels in the samples with a $b$ tag are 
summarized in \Tab{t:backgroundnumber}.
The background level in the 0-tag sample is determined in the subsequent fit.

\begin{table}
\caption[Sample composition.]
{The background composition and the number of \ttbar
candidates for events with $\geq1$ $b$ tag, and
for the subset used in the DLM analysis.  Statistical and systematic
uncertainties have been combined.}
\label{t:backgroundnumber}
\begin{ruledtabular}
\begin{tabular}{lcc}
                         & $\geq1$ $b$ tag & DLM sample \\
\hline
Source & \multicolumn{2}{c}{Expected Background} \\
\cline{2-3}
$W$ + jets          & \measErr{19.6}{2.4} & \measErr{5.3}{1.1} \\
Multijet            & \measErr{4.7}{0.7} & \measErr{3.1}{1.0} \\
Other               & \measErr{2.3}{0.2} & \measErr{0.8}{0.1} \\
\hline
Total               & \measErr{26.6}{3.0} & \measErr{9.2}{1.8} \\
\hline\hline
& \multicolumn{2}{c}{Selected \ttbar Candidates} \\
\cline{2-3}
Data                &  121 & 63 \\
\end{tabular}
\end{ruledtabular}
\end{table}

Both analyses use calibrated jet energies, based on
a combination of instrumental calibration and
analysis of data control samples~\cite{r_NIMJES}.
The uncertainty $\sigma_c$\ on the jet energy scale 
is the a major source of uncertainty on
\mtop. For jets in the \ttbar sample, $\sigma_c$ is approximately
3\% of the measured jet energy, depending on the $\eta$ and \pt of the jet.
The parameter \jes is defined as the difference, averaged over all jets
in the sample, between the true jet energy and our measured jet energy
after calibration in units of $\sigma_c$.

%
%
%
%

The DLM technique, described in detail in \cite{r_Run2DLM}, defines a
likelihood for each event based on the
differential cross section per unit phase space volume of the final state
partons, $d\sigma_{\ttbar}/d\Phi$, 
as a function of \mtop.
Detector resolution effects are accounted for using \ttbar\ events 
generated by the HERWIG Monte Carlo program \cite{r_herwig}\ and full 
detector simulation to derive a transfer function (TF). 
The TF relates the transverse energies of the quarks,
denoted by $\bm{x}$, and the observed jets.
For a given event, a Monte Carlo integration is performed over the
possible \ttbar\ final state kinematics in the following way:  
we first
generate a random value for the virtual mass squared of the $W$ boson
in the leptonic channel, $s_{W}$, according to the Breit-Wigner form.
We identify the momentum of the electron or muon daughter with the measured
value, and the neutrino transverse momentum with the measured
missing transverse energy.  We then
generate random values for the momenta of final state quarks
according to the TF probabilities.
We determine the $z$ component of the neutrino momentum, with a two-fold
ambiguity, using $s_W$ as a constraint.
Thus, for a given set of $\bm{x}$ and $s_{W}$, we fully determine
the event kinematics, and the event likelihood as a function of \mtop is given by
\begin{equation}
L(\mtop) = N \sum_{I_{j}} \sum_{I_{\nu}}
                 \frac{d\sigma_{\ttbar}}{d\Phi}
                 (\mtop; \bm{x}, s_W),
\end{equation}
where the normalization factor $N$\ is independent of \mtop for a given
event, and the indices $I_j$ and $I_{\nu}$ run over the parton-jet
assignments and the two neutrino solutions, respectively.
The event likelihood is obtained by numerically integrating
over $\bm{x}$ given by the TF and $s_W$ given by the Breit-Wigner
distribution.

\Fig{f:DLMMaxLikelihood} shows the distribution of the top quark
mass value at the point of maximum likelihood in each event
compared with the expectation from simulated events.
An inset shows the joint log-likelihood as a function of \mtop,
formed by multiplying the likelihoods of the individual events together.
We account for the presence of background events by evaluating
the shift of $-\gevcc{1.4}$\ they make in the measured top quark mass. 
From the joint likelihood we infer 
$\mtop = \gevcc{\measAStat{173.2}{2.6}{2.4}}$,
where the uncertainty is only statistical.
The systematic uncertainty due to the jet energy scale is estimated
as the shift in \mtop arising from a \sigcunit{1} change in \jes,
and is \gevcc{3.0}.

\begin{cfigure}
\includegraphics{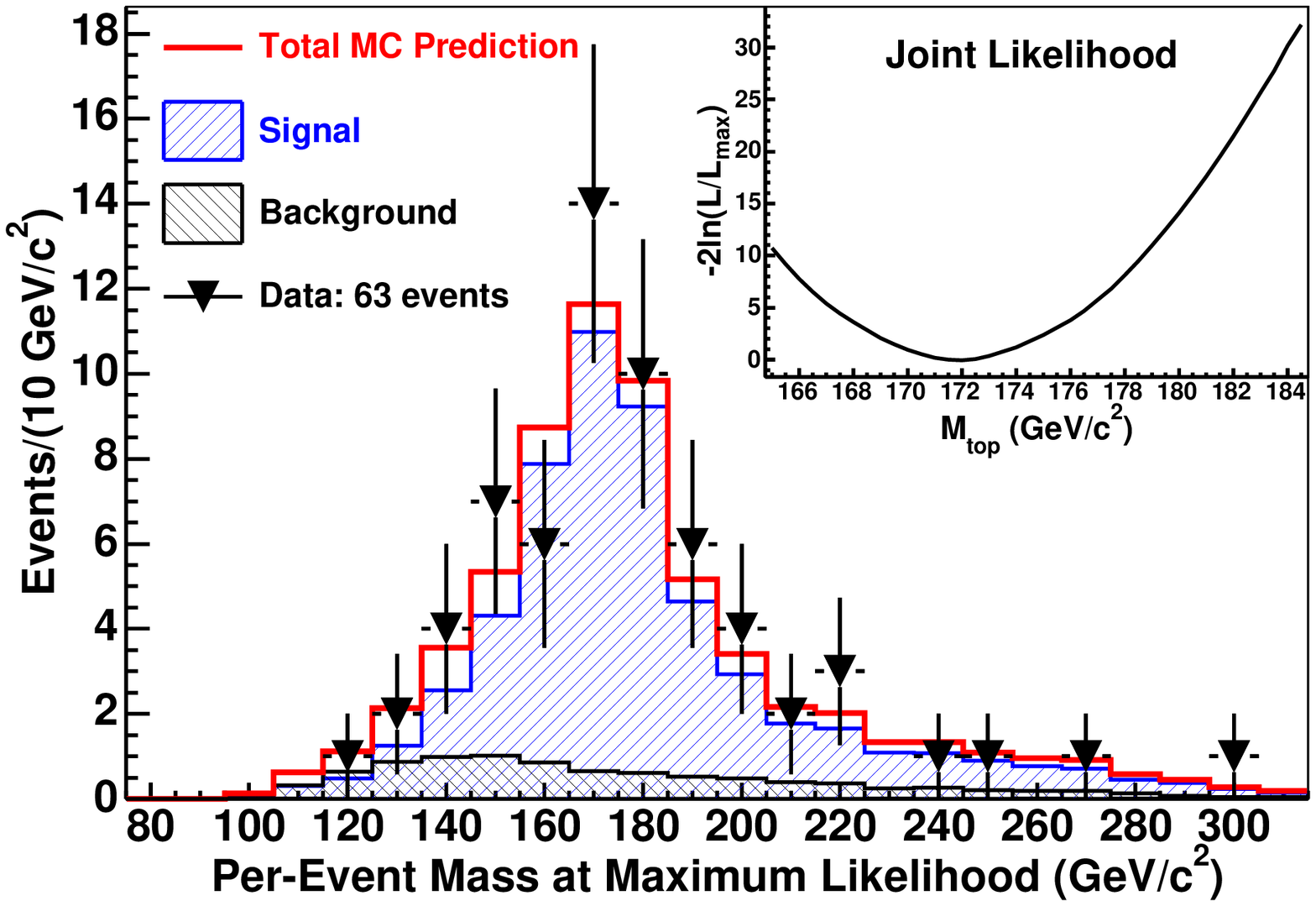}
\caption[DLM max-likelihood.]
{The value of the top quark mass at the maximum of the DLM
likelihood is plotted for each event. Data events (points)
are compared
to an expected distribution (histogram) comprising simulated \ttbar
(\mtop=\gevcc{172.5}) and background events.  The last bin includes
events with masses $>\gevcc{305}$.
The inset shows the 
joint log-likelihood for the 63 events, before accounting for the presence
of background.}
\label{f:DLMMaxLikelihood}
\end{cfigure}

%
%
%
%

The template method is described in detail in~\cite{r_Run2TMT}.
We perform a \chisq minimization to fit the parton momenta from the \ttbar
daughters and determine \mreco for each event,
assuming that the final state arises from the decay of a 
\ttbar pair into $W$ bosons and $b$ quarks.
We use only the four leading jets in the mass reconstruction.
In the \chisq fit, both sets of $W$ decay daughters are constrained 
to have the invariant mass of the $W$ boson, and both $Wb$ states are
constrained to have the same mass.
The ambiguity arising from the different ways of assigning the 
jets to the four quarks
is resolved by selecting the assignment with the lowest \chisq,
taking into account the $b$-tagging information.
We construct a histogram of \mreco for each subsample, discarding
events with \chisq$>9$, corresponding to poorly reconstructed
or background events.

The parameter \jes\ is determined within this event sample by
removing the $W$ boson mass constraints, and identifying for each
event all pairs of jets that would be consistent
with the $W$ boson final state.
We form histograms of the invariant masses of these jet pairs for
each of the four event subsamples and compare these with what we expect 
given the precisely known $W$ boson mass~\cite{r_PDBook}.

We use these eight histograms to measure simultaneously \mtop\ and \jes. 
An unbinned likelihood fit is performed to
parameterized signal templates taken from simulated \ttbar events generated using
different values of \mtop and \jes, and background templates derived
from studies of the relevant background processes. 
We include in the fit a Gaussian constraint ($\jes=\sigcunit{\measErr{0}{1}}$)
from the extrinsic jet energy calibrations,
and we constrain the background rates in the 2-tag, 1-tag Tight,
and 1-tag Loose samples to the estimated background rates within
their uncertainties.
The background level in the 0-tag sample is determined to be
$\measAStat{15.7}{8.0}{7.1}$\ events by the fit using
the differences in predicted signal and background mass distributions. 

The four reconstructed top quark mass distributions and the results of
the fit are shown in \fig{f:fitConsistencyMreco},
where we also show the background contributions. 
In all cases, we see agreement between the observed data
distributions and the fitted curves. We obtain 
$\mtop = \gevcc{\measAStat{173.5}{3.7}{3.6}}$,
where the uncertainty is statistical and incorporates the
uncertainty due to \jes, which we estimate contributes
$\sim\gevcc{2.5}$. \Fig{f:lnlcomb} shows the
likelihood in the \mtop-\jes plane. 
If we do not constrain \jes to the nominal value of zero, we obtain 
$\jes = \sigcunit{\measErr{-0.25}{1.22}}$, which indicates 
our nominal jet energy calibrations are in good agreement with 
information provided by the $W$ boson mass peak in the \ttbar decay.
This also demonstrates that the \jes\ constraint from the $W$\ boson decay
has comparable precision to the jet energy calibration.

\begin{cfigure}
\includegraphics{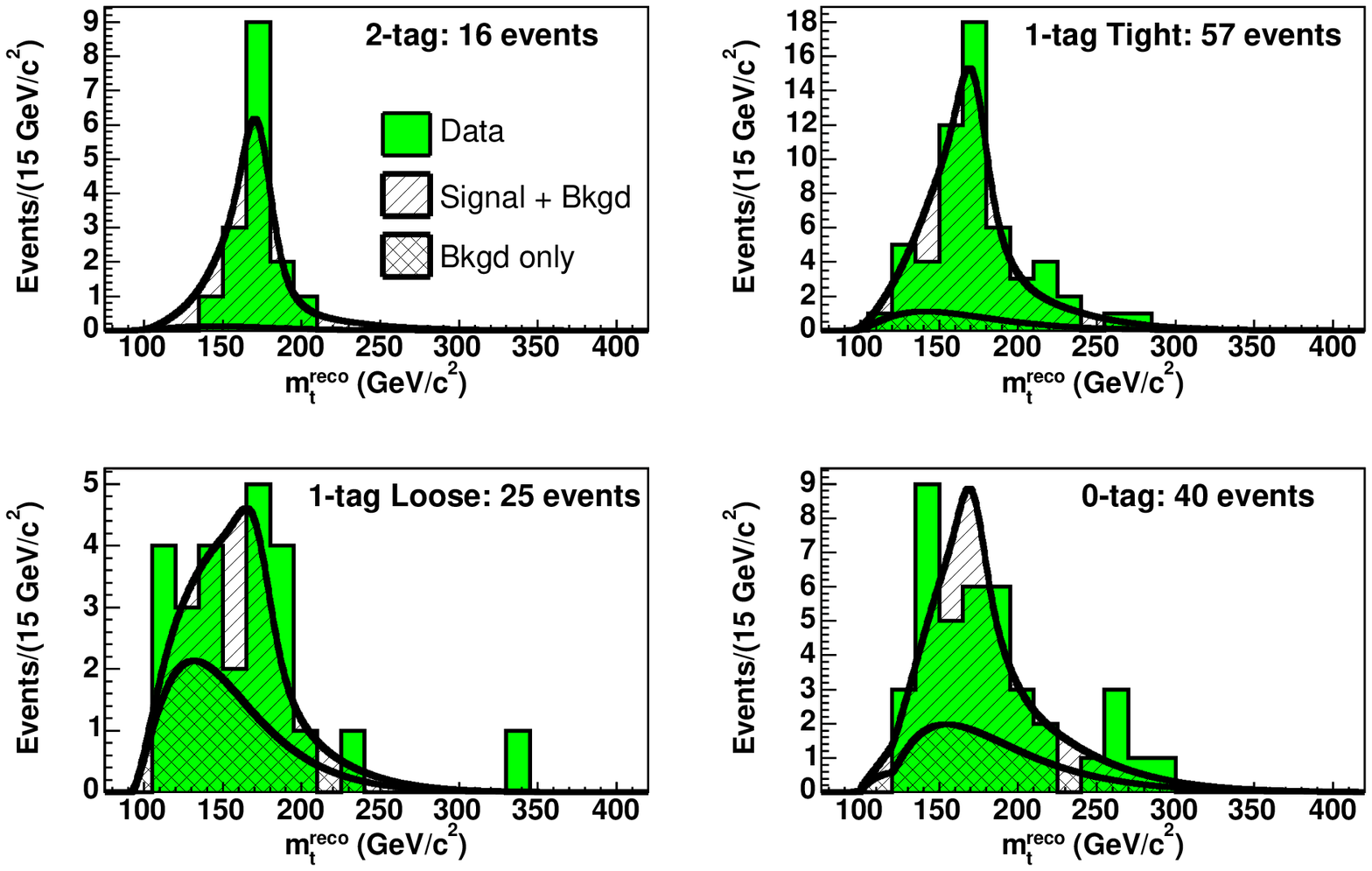}
\caption[Subsample fits consistency checks: \mreco.]
{The \mreco distribution of the template method is shown
for each subsample overlaid with the expected distribution using \mtop,
\jes and signal and background
normalizations from the combined fit.  The events with $\chi^2>9$\ have
been excluded from each subsample.}
\label{f:fitConsistencyMreco}
\end{cfigure}

\begin{cfigure}
\includegraphics{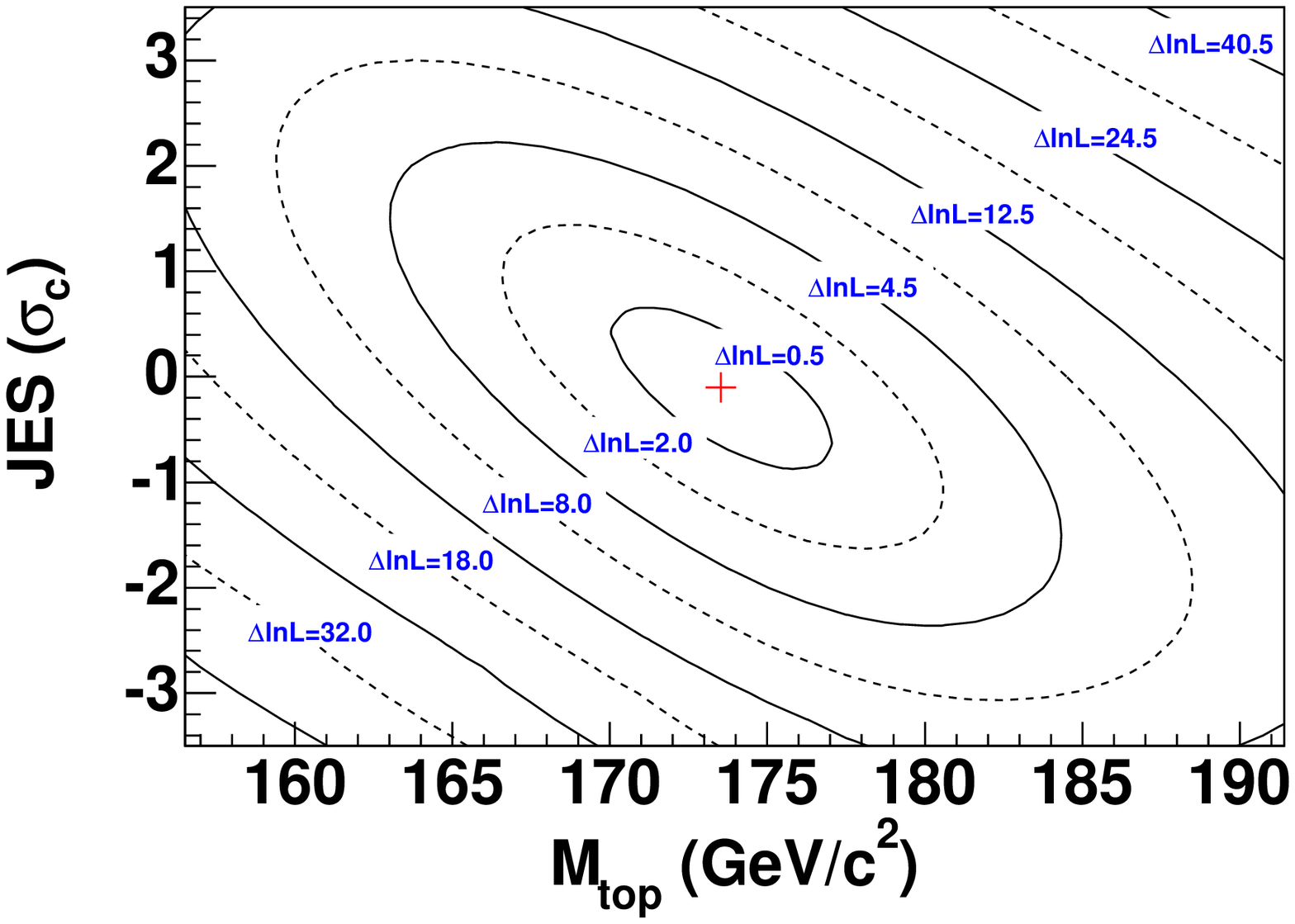}
\caption[Combined fit likelihood contours.]
{Contours of the template method likelihood are shown in the
\mtop-\jes plane for the combined fit to all four subsamples.
The crosshair shows the best fit point. Contours are
given at intervals of $\Delta\ln L$, the change in
log-likelihood from its maximum.}
\label{f:lnlcomb}
\end{cfigure}

%
%
%
%

There are a number of additional systematic uncertainties that 
affect both analyses:
initial state and final state radiation uncertainties (ISR/FSR),
uncertainties arising from the parton distribution functions (PDFs),
and uncertainties arising from 
modeling of the background processes, 
the choice of event generators and 
$b$-jet fragmentation, decays, and
color connections (Modeling) \cite{r_pythia,r_alpgen}.
\Tab{t:uncertainties} summarizes these uncertainties. 

\begin{table}
\caption[Systematic uncertainty summary.]
{The systematic uncertainties for the
two analyses.}
\label{t:uncertainties}
\begin{ruledtabular}
\begin{tabular}{ldd}
Systematic &
\multicolumn{1}{c}{DLM} &
\multicolumn{1}{c}{Template} \\
&
\multicolumn{1}{c}{$\Delta\mtop$ (\gevccnoarg)} &
\multicolumn{1}{c}{$\Delta\mtop$ (\gevccnoarg)} \\
\hline
Jet Energy Scale& 3.0 &
\multicolumn{1}{c}{$[\sim 2.5]$\footnotemark[1]} \\
ISR/FSR         & 0.6 & 0.7 \\
PDFs            & 0.5 & 0.3 \\
Modeling        & 0.7 & 0.9 \\
Method          & 0.5 & 0.6 \\
\hline
Total 			    & 3.2 & 1.3\footnotemark[1]
\footnotetext[1]{\label{f:systtable}The \jes uncertainty
is included in the uncertainty reported by the likelihood fit.}
\end{tabular}
\end{ruledtabular}
\end{table}

%
The DLM method has additional uncertainties that arise from the
use of transfer functions and from the procedure that corrects the
measured mass for the presence of background (Method).
Together with the \jes and other common sources noted above, the systematic
uncertainty on the DLM mass measurement is \gevcc{3.2}.

%
The template method has additional uncertainties arising from the 
statistical precision of the templates themselves and approximations
made in treating \jes as a single parameter affecting 
all jets coherently (Method).
The total systematic uncertainty on the
template mass measurement is \gevcc{1.3}.

%
%
%
%

In summary, we have presented two new measurements of the top quark mass.
The analysis using the DLM method results in 
$\mtop = \gevcc{\measAStatSyst{173.2}{2.6}{2.4}{3.2}}$;
the analysis using the template technique results in
$\mtop = \gevcc{\measAStatSyst{173.5}{3.7}{3.6}{1.3}}$. 
There is a large statistical correlation between these measurements
given the common data sample,
so that we quote as a result only the more accurate
measurement, the template method result of
$\mtop = \gevcc{\measAErr{173.5}{3.9}{3.8}}$.
This provides the most precise single measurement on this 
important physical parameter. 
In comparison, the previous most precise measurement was
$\mtop = \gevcc{\measErr{180.1}{5.3}}$\ \cite{Abazov:2004cs} 
and the world average was 
$\mtop = \gevcc{\measErr{174.3}{5.1}}$\ \cite{r_PDBook}.

\begin{acknowledgments}
We thank the Fermilab staff and the technical staffs of the participating
institutions for their vital contributions. This work was supported by the
U.S. Department of Energy and National Science Foundation; the Italian
Istituto Nazionale di Fisica Nucleare; the Ministry of Education, Culture,
Sports, Science and Technology of Japan; the Natural Sciences and
Engineering Research Council of Canada; the National Science Council of
the Republic of China; the Swiss National Science Foundation; the A.P.
Sloan Foundation; the Bundesministerium f\"ur Bildung und Forschung,
Germany; the Korean Science and Engineering Foundation and the Korean
Research Foundation; the Particle Physics and Astronomy Research Council
and the Royal Society, UK; the Russian Foundation for Basic Research;
the Comisi\'on Interministerial de Ciencia y Tecnolog\'{\i}a, Spain;
in part by the European Community's Human Potential Programme under
contract HPRN-CT-2002-00292; and the Academy of Finland.
\end{acknowledgments}

\bibliographystyle{apsrev}
\bibliography{topMassPRL}

\end{document}